\documentclass[english]{lni}
\usepackage{braket}
\usepackage{mathtools}

\begin{document}
%%
%%% Mehrere Autoren werden durch \and voneinander getrennt.
%%% Die Fußnote enthält die Adresse sowie eine E-Mail-Adresse.
%%% Das optionale Argument (sofern angegeben) wird für die Kopfzeile verwendet.
\title[MDE4QAI]{Towards Model-Driven Engineering for Quantum AI}
%%%\subtitle{Untertitel / Subtitle} % if needed
\author[]
{Armin Moin\footnote{Technical University of Munich, University of Antwerp \& Flanders Make, Germany \& Belgium. \email{armin.moin@tum.de}} \and
Moharram Challenger\footnote{University of Antwerp \& Flanders Make, Belgium. \email{moharram.challenger@uantwerpen.be}} \and
Atta Badii\footnote{University of Reading, United Kingdom.
\email{atta.badii@reading.ac.uk}} \and
Stephan Günnemann\footnote{Technical University of Munich \& Munich Data Science Institute (MDSI), Germany. \email{guennemann@in.tum.de}}
}
\startpage{1} % Beginn der Seitenzählung für diesen Beitrag / Start page
\editor{Herausgeber et al.} % Names of Editors
\booktitle{GI QC 22: GI Quantum Computing Workshop} % Name of book title
%\yearofpublication{2022}
%%%\lnidoi{18.18420/provided-by-editor-02} % if known
\maketitle

\begin{abstract}
Over the past decade, Artificial Intelligence (AI) has provided enormous new possibilities and opportunities, but also new demands and requirements for software systems. In particular, Machine Learning (ML) has proven useful in almost every vertical application domain. In the decade ahead, an unprecedented paradigm shift from classical computing towards Quantum Computing (QC), with perhaps a quantum-classical hybrid model, is expected. We argue that the Model-Driven Engineering (MDE) paradigm can be an enabler and a facilitator, when it comes to the quantum and the quantum-classical hybrid applications. This includes not only automated code generation, but also automated model checking and verification, as well as model analysis in the early design phases, and model-to-model transformations both at the design-time and at the runtime. In this paper, the vision is focused on MDE for Quantum AI, particularly Quantum ML for the Internet of Things (IoT) and smart Cyber-Physical Systems (CPS) applications.
\end{abstract}
%%% Beginn des Artikeltexts

	\section{Introduction}\label{introduction}
		In October 2019, Arute et al. \cite{Arute+2019} published the breakthrough results of their research and development at Google AI Quantum on \textit{Quantum Computing (QC)}, where they reported the total runtime of only 200 seconds for a specific computational task on their programmable superconducting quantum processor, which would take approximately 10,000 years on a state-of-the-art \textit{classical} (i.e., non-quantum) supercomputer. This groundbreaking experiment validated the previously hypothetical and theoretical vision of \textit{quantum supremacy}, initiated by Richard Feynman in the 1980s. Today, it is a realistic expectation that QC will eventually be a revolutionary and disruptive enabler technology in the upcoming years and decades in various domains and disciplines: from cryptography and security to Artificial Intelligence (AI), many pieces of software and technologies will become obsolete. Consequently, the current practitioners, such as software developers and data scientists would - in principle - need to learn the new programming paradigms, APIs, and frameworks for QC, in order to be able to deal with the new generation of computers, namely quantum computers. However, current software applications will not abruptly disappear or become useless. It is very likely that for a relatively long period of time, we will have to deal with a mixture of classical and quantum computers, algorithms, and data. Hence, the current challenges introduced by pervasive, distributed computing, which intrinsically involve heterogeneity, will even become more crucial, due to the new hardware technologies and architectures, which will require a fundamentally different paradigm and model for the programming and execution.
		
		Based on our experiences with the heterogeneous and distributed Internet of Things (IoT) services and the highly complex systems of systems, namely the smart Cyber-Physical Systems (CPS), as well as AI \cite{ML-Quadrat, Moin+2018, Moin+2020}, where we found the applications of the Model-Driven Software Engineering (MDSE) paradigm, particularly the Domain-Specific Modeling (DSM) methodology \cite{KellyTolvanen2008} with full code generation beneficial, we maintain the idea of enabling MDSE for quantum computers and hybrid applications, that shall run on a mixture of quantum and classical computers. The overall idea of modeling quantum programs has been recently proposed independently by Delgado and Gonzalez \cite{DelgadoGonzalez2020},	Ali and Yue \cite{AliYue2020}, as well as Gemeinhardt et al. \cite{Gemeinhardt+2021}.
		
		The contribution of this paper is proposing our novel vision towards MDE4QAI, where our current open source technology stack concerning domain-specific MDSE for smart CPS and the IoT \cite{ML-Quadrat} should be extended to support QAI, including Quantum ML (QML) and Quantum Multi-Agent Systems (QMAS). This inherently includes the hybrid quantum-classical applications as well.
		
		The rest of this paper is structured as follows: Section \ref{sota} briefly reviews the state of the art. Further, we propose our vision concerning MDE4QAI with a focus on smart CPS and the IoT in Section \ref{vision}. Finally, we conclude and suggest the future work in Section \ref{conclusion-future-work}.
	
	\section{State of the Art}\label{sota}
		Today's non-quantum (i.e., \textit{classical}) computers deal with bits. However, built based on quantum mechanics, quantum computers deal with quantum bits, known as \textit{qubits}. A qubit may simultaneously represent a 0 and a 1, with certain degrees or probabilities of 0-ness and 1-ness. This property is called \textit{superposition}. For example, if we have 2 bits, they can encode 4 possible states, namely $00$, $01$, $10$, and $11$, whereas 2 qubits can be represented through the following linear combination in the bra-ket (Dirac) notation: $\alpha_0 \ket{00} + \alpha_1 \ket{01} + \alpha_2 \ket{10} + \alpha_3 \ket{11}$, where $\alpha_i$ are complex numbers, called amplitudes, and their squares (i.e., $(\alpha_i)^2$) sum to 1. In fact, $(\alpha_i)^2$ is the probability that the quantum program will be in the state that is associated to the amplitude $\alpha_i$, once the qubits are read by the quantum computer. In addition to superposition, there is another characteristic of quantum mechanics, thus quantum information/computation, called \textit{entanglement}. If a pair or a group of quantum particles or qubits in our context are entangled, it is not possible to describe the quantum state of one particle/qubit independently of the other one(s). Therefore, the outcome of the measurement for one of the entangled particles/qubits is always correlated with the outcome of the measurement for the other one(s) even if they are far away \cite{AliYue2020, NielsenChuang2011}.
		
		Quantum-encoded data require incomparably less storage space. Just as multiple bits can represent exponentially many numbers, a multi-qubit system can represent a superposition of exponentially many bit strings. \textit{Quantum data} construct one pillar of QC, thus, one pillar of QAI/QML \cite{Arens2018}. The second pillar is shaped by \textit{quantum algorithms}. Note that any classical algorithm can be executed on quantum computers too. However, the QC community aims for exploiting the unique quantum properties, such as entanglement, in algorithms, to enable performance leap. The algorithms that inherently benefit from the essential properties of QC are called quantum algorithms. Nevertheless, the unsolvable computational problems (in terms of computability) on classical computers remain unsolvable on quantum computers too. Only the solvable problems could be solved much more efficiently on quantum processors. Examples include, but are not limited to the Shor’s Algorithm for integer factorization and discrete logarithms \cite{Shor1994}, the Grover's algorithm for efficient database search \cite{Grover1996}, graph algorithms and random walks, as well as Principle Component Analysis (PCA) \cite{LANL2018}.
	
	    There exist various models of computation for QC, such as quantum circuit/logic gate, adiabatic/annealing, topological, quantum walks, one clean qubit, measurement-based and quantum Turing machines \cite{Jordan2008}. In practice, the quantum circuit model is well established, whereas quantum annealing is also deployed in certain implementations, for example, the Quantum Processing Units (QPU) of D-Wave Systems \cite{DWaveSystems}. In fact, it is already proven that any quantum circuit algorithm can be transformed into a quantum annealing algorithm with the exact same time complexity and vice-versa, hence showing they are essentially equivalent \cite{Yu+2018, Johnston2018}. Furthermore, the key technologies for the actual physical implementation of quantum processors comprise superconducting, ultra-cold atoms (e.g., trapped-ions), quantum optics and spin-based. For instance, Linke et al. \cite{Linke+2017} conducted experiments on two 5-qubit quantum computers with the quantum circuit model, which had two different technologies/architectures, namely superconducting and trapped-ions. They illustrated that the superconducting system offered faster gate clock speeds and a solid-state platform, whereas the other one featured superior qubits and reconfigurable connections. They concluded that the performance of those systems reflected the topology of connections in the base hardware, thus supporting the idea that quantum software and hardware should be ideally co-designed \cite{Linke+2017}. 
		
		IBM \cite{IBM-Q} adopted the quantum circuit model of computation and the superconducting technology. Analogue to the low-level Assembly programming languages for classical computers, the OpenQASM \cite{Cross+2017, OpenQASM} provided an intermediate representation for the quantum instructions of the IBM quantum computers. A higher level API is provided by the open source SDK, Qiskit \cite{Qiskit}. The programmer may either create the circuits in Python through Jupyter notebooks or via a GUI. Additionally, Microsoft provided an open source SDK, called Q\# \cite{Svore+2018}, which can be used with their Azure Quantum open cloud ecosystem \cite{AzureQuantum}. They enabled access to diverse quantum hardware technologies of their partners, ranging from trapped-ions (Honeywell/IONQ) to superconducting (Quantum Circuits, Inc.). Further, D-Wave Systems \cite{DWaveSystems} offered the open source Ocean SDK \cite{ocean-sdk} for using their superconducting quantum annealing QPUs. Last but not least, the Google quantum supremacy experiment \cite{Arute+2019} (see Section \ref{introduction}) was conducted on a superconducting quantum computer, specifically a Noisy Intermediate-Scale Quantum (NISQ) \cite{Preskill2018} processor with 53 qubits \cite{TFQ-Concepts}. Most of the current gate model solutions are NISQ.
		
		Google/Alphabet, who provided TenforFlow \cite{Abadi+2015}, an open source framework and Python library for ML using Artificial Neural Networks (ANN), particularly deep learning, are also one of the pioneers of QAI/QML. They extended TensorFlow for hybrid quantum-classical ML. This extended open source library is called TensorFlow Quantum (TFQ) \cite{TFQ-Concepts}. It deploys Cirq \cite{Cirq}, an open source Python library for writing, manipulating, and optimizing quantum circuits and running them against the NISQ quantum computers and simulators. Cirq is the alternative solution of Google to Qiskit \cite{Qiskit}, Q\# \cite{Svore+2018} and Ocean \cite{ocean-sdk}. The latter also offers special support for ML (e.g., for probabilistic models). However, QAI is not limited to QML. Already in 2004, Klusch \cite{Klusch2004a, Klusch2004b} proposed the perspective of using Intelligent Agents (IA) and Multi-Agent Systems (MAS) in QC. Other works in the IA/MAS community (e.g., Chen et al. \cite{Chen+2007}) also studied QC or specific properties of it. However, real-world implementations of IA/MAS on quantum computers, or more precisely hybrid quantum-classical MAS are emerging only recently, for example, see \cite{Neumann2020, Kirke2020}.
	
	\section{Proposed Vision: MDE4QAI}\label{vision}
		In this work, we focus on MDE in the context of Software Engineering (SE), thus MDSE. We first draw our proposed vision regarding MDE4QC and then concentrate specifically on MDE4QAI. The MDSE paradigm, which provides abstraction and automation can hide the complexity and increase the productivity of the software development. There exist many benefits in deploying software models as the central artifacts in the Software Development Life-Cycle (SDLC), for example, concerning the early validation, verification, analysis and simulation, as well as the automated generation of the implementation artifacts, such as the source code and the documentation. In our prior work, ML-Quadrat \cite{Moin+2018, Moin+2020, ML-Quadrat}, which was based on ThingML/HEADS \cite{ThingML,Harrand+2016,HEADS}, we deployed the DSM methodology \cite{KellyTolvanen2008} of MDSE with full code generation to support the development of smart IoT services and smart CPS applications that needed to deploy ML. Thus, we provided a unified layer of abstraction to the practitioners, namely the users of the Domain-Specific Modeling Language (DSML), despite all of the heterogeneity of the underlying IoT platforms and the diversity of the ML libraries, frameworks and methods. Further, other examples in the literature include, but, are not limited to PIM4Agents \cite{WarwasKlusch2011}, that abstracted from various MAS platforms (e.g., JACK, JADE, and Jadex), and the work of Challenger et al. \cite{Challenger+2014}. Since the Agent-Oriented Software Engineering (AOSE) paradigm fits very well to the IoT/CPS domain (see \cite{PicoHolgado2018, GeisbergerBroy2014}), the integration of the approaches as mentioned above (e.g., ML-Quadrat and PIM4Agents) constitutes the first part of the proposed vision in this paper. Note that the prior work on which ML-Quadart was based \cite{Moin+2018, Moin+2020, ML-Quadrat}, namely ThingML/HEADS \cite{ThingML,Harrand+2016,HEADS}, did not support AI at all. However, ML-Quadrat enabled ML, which was the first step towards enabling AI and cognitive capabilities -- which are vital for the smart services in the IoT-related domains -- in the modeling framework of ThingML. However, extending this AI-support towards automated reasoning and intelligent agents will be the next necessary step in this direction.
		
\begin{figure}[!t]
			\centerline{\includegraphics[width=0.55\textwidth]{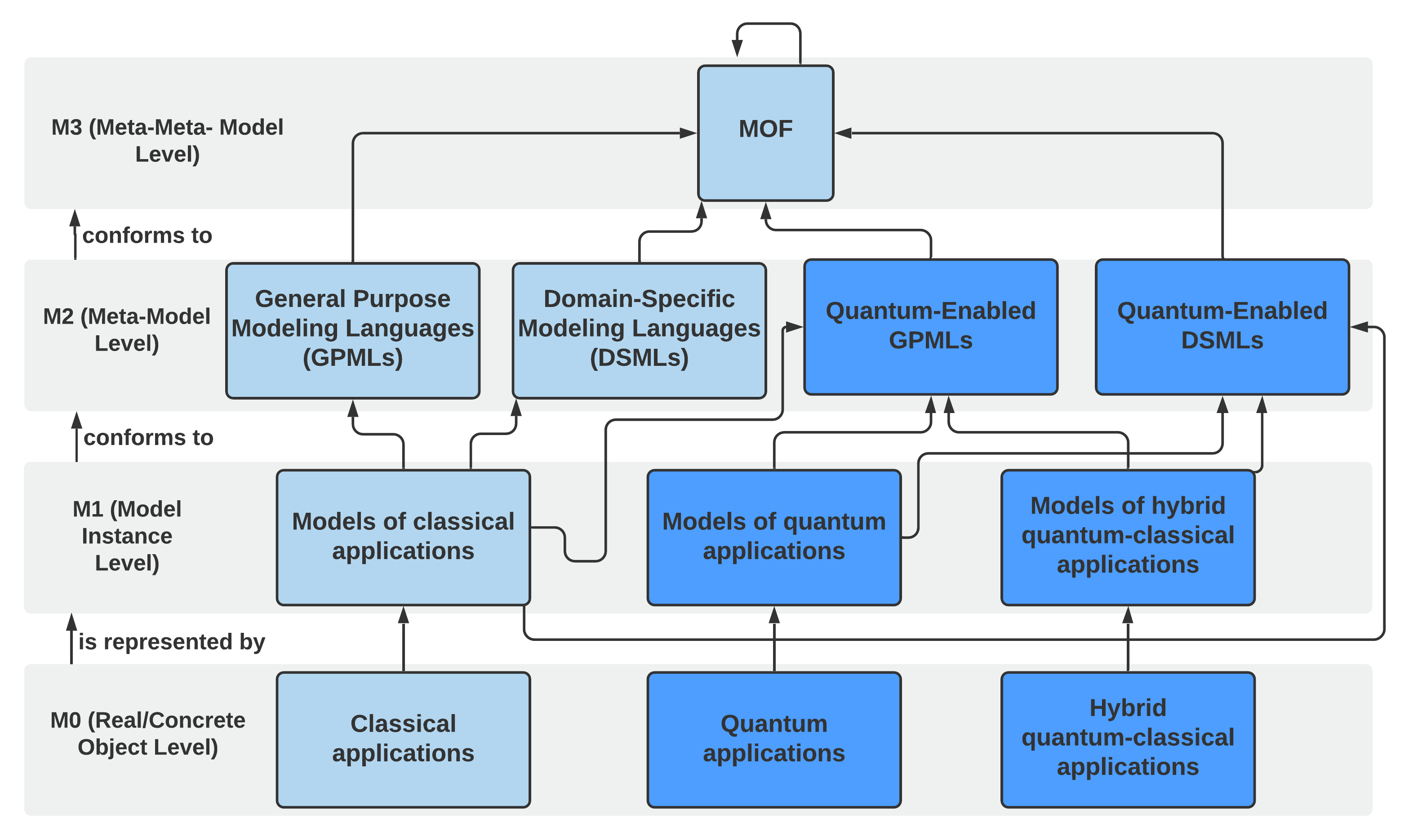}}
			\caption{The proposed vision of MDE4QC in the context of MOF-based meta-modeling.}
			\label{fig:mde4qc}
		\end{figure}

		In addition, the second part of the proposed vision is enabling classical modeling languages such that they can support the modeling of hybrid quantum-classical software applications. Figure \ref{fig:mde4qc} illustrates this idea in the context of the Object Management Group (OMG) and ISO/IEC standard for meta-models, called Meta-Object Facility (MOF) \cite{OMGMOF,ISOMOF}. While we share the overall idea proposed by Gemeinhardt et al. \cite{Gemeinhardt+2021}, we advocate the extension of the existing modeling languages, both the General-Purpose Modeling Languages (GPMLs), such as UML/SysML and the DSMLs, such as ML-Quadrat \cite{Moin+2018, Moin+2020, ML-Quadrat} and PIM4Agents \cite{WarwasKlusch2011}. This vision is in line with the approach of Delgado and Gonzalez \cite{DelgadoGonzalez2020}, who proposed a UML extension for QC, called Q-UML. In contrast, Gemeinhardt et al. \cite{Gemeinhardt+2021} considered a separate category for quantum modeling languages, next to GPMLs (UML) and DSMLs. However, we believe that no matter how far the programming paradigms, APIs and technologies of the classical and quantum computers are, modeling languages and DSMLs of the future shall continue to offer sufficiently extendable and adaptive models to support the requisite abstractions for the pure classical/quantum and the hybrid quantum-classical software systems. This is because we expect the co-existence of classical and quantum processors with various architectures and technologies in the foreseeable future.
		
		Stepping into the world of quantum Information and Communication Technologies (ICT) and the hybrid quantum-classical ICT world, we shall expect not only more distributed computation, but also much more heterogeneity. This may comprise quantum/hybrid vs. classical data/information, quantum/hybrid vs. classical algorithms/computation, and quantum vs. classical communication methods and technologies. Hence, a mixture of classical and quantum processors, each with different architectures and computational powers, as well as their individual constraints (e.g., in terms of the electric power consumption or the environmental requirements, such as, concerning the temperature, noise, and possible signal interference) for the coming years and possibly decades will be likely. In this setup, MDSE shall be capable of providing even more added value than before, given the unprecedented heterogeneity of the target hardware and software platforms, as well as the enormous complexity of the counter-intuitive and alien world of QC. MDSE will be capable of abstracting from the low-level APIs and programming paradigms, thus offering a higher level of abstraction, where practitioners can model their target applications, regardless of whether the applications shall run fully or partially on quantum processors of potentially diverse technologies and architectures. The advantages of MDSE and DSM may go far beyond the automated generation of the implementation artifacts. Model analysis for program comprehension and model checking for formal verification in the early design phases, which are typically interesting in the classical systems will be even more vital for QC and hybrid applications, due to the abundance of heterogeneous and distributed hardware and software platforms, the unfamiliarity of the technologies, as well as the limited access to the QC resources. The latter is currently addressed through the QC simulators on classical computers, but for the real-world distributed use-case scenarios with the hybrid quantum-classical applications, this does not suffice. However, the said vision implies the formalization and realization of the QC domain semantics in general, as well as the semantics and constraints of the specific target platforms for QC, ideally directly on the meta-model layer, but also possibly through the model-to-model and the model-to-code transformations. In other words, the domain knowledge regarding QC must be transferred from the human experts and the available libraries (mentioned in Section \ref{sota}) to the modeling languages and tools.
		
\begin{figure*}[!t]
			\centerline{\includegraphics[width=1.0\textwidth]{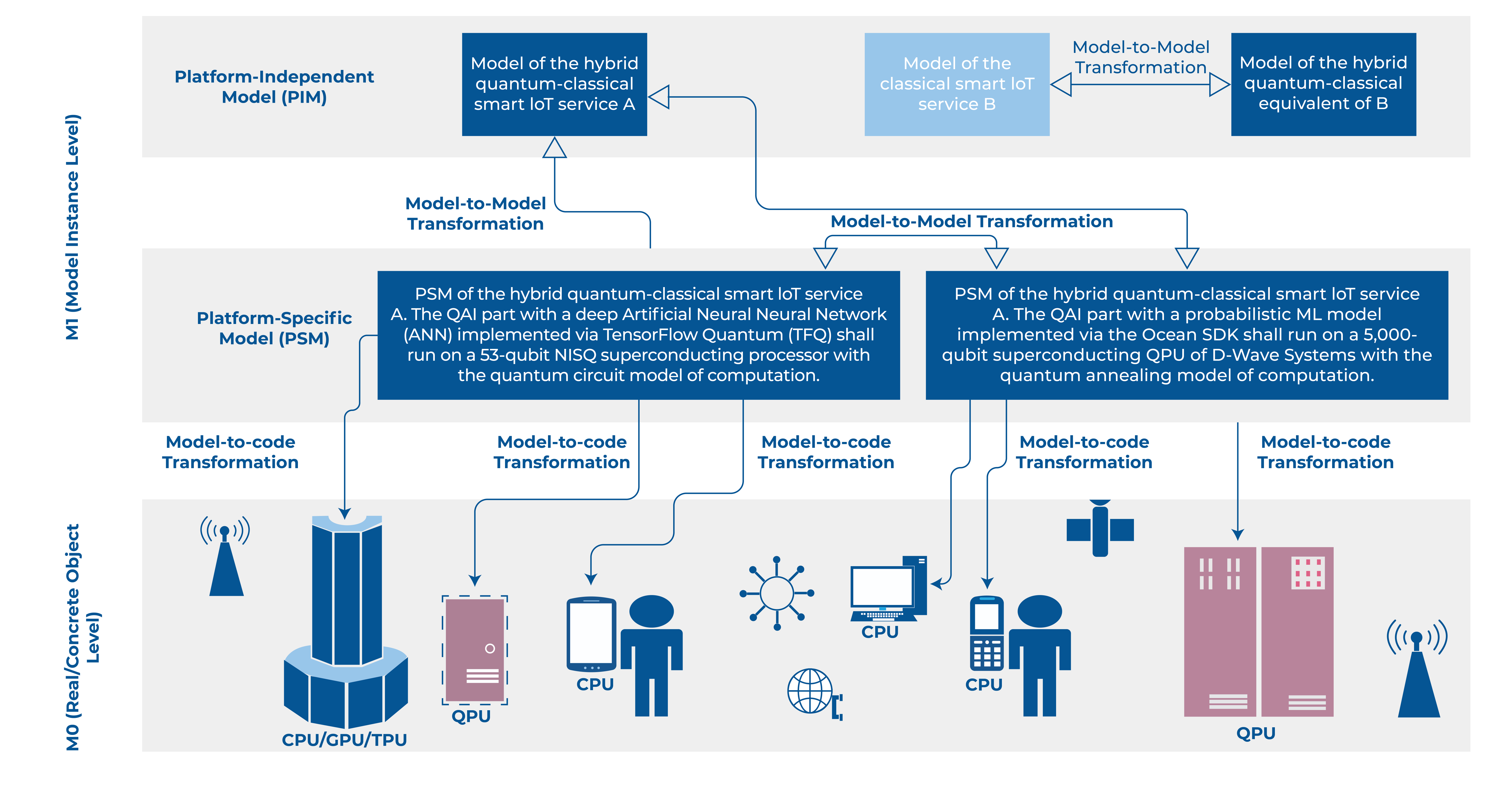}}
			\caption{The proposed vision of MDE4QAI in the context of Model-Driven Architecture.}
			\label{fig:mde4qai}
		\end{figure*}

		Since our focus is on MDE4QAI, we plan to extend our existing DSML and modeling tool, ML-Quadrat towards QAI, specifically Quantum ML (QML) and Quantum MAS (QMAS). Currently, the practitioner using ML-Quadrat or similar domain-specific MDSE tools, does not need to be concerned about the technical details of the underlying platforms and their APIs or programming paradigms. In fact, in some cases, the practitioner might not even know at the design-time which specific IoT hardware/software platforms or communication protocols will be used at the run-time. Similarly, they might not have any idea at the design-time whether the ML algorithms will run on CPUs, GPUs, Tensor Processing Unit (TPUs) or a combination of them. Many such decisions can be left open to be made at the run-time, or can be set according to certain criteria. What is important is to offer the practitioner a higher layer of abstraction (i.e., the modeling level), where they can focus on the business logic without being concerned about the underlying technologies, which might change at a fast pace or become (un)available at short notice (see the Models@Runtime approaches, such as \cite{Fouquet+2012}). By expanding the range of supported platforms and technologies to QC, we will consider not only CPUs, GPUs and TPUs, but also various QPUs with different models of computation, architectures, technologies, hence diverse powers and constraints. The envisioned scenario for deploying MDE for QAI (MDE4QAI) in the context of Model-Driven Architecture (MDA) \cite{OMGMDA2014} is illustrated in Figure \ref{fig:mde4qai}. Model-to-code transformations shall enable the fully-automated code generation. Moreover, it would be interesting to examine how far model-to-model transformations might be able to facilitate at the modeling level the following transformations: (i) A purely classical application to its hybrid/quantum equivalent or vice-versa, or (ii) a purely quantum application to its hybrid/classical equivalent or vice-versa, or (iii) a quantum/hybrid application designed for a certain model of computation, say quantum circuits to its equivalent designed for another model of computation, such as quantum annealing or vice-versa, or (iv) a quantum/hybrid application tuned for running on a specific quantum processor with a particular architecture and technology, say trapped-ions with X qubits to its equivalent optimized for running on another quantum processor with a different architecture and technology, such as a superconducting NISQ with Y qubits. Last but not least, as mentioned, analysis, verification and simulation in the early design stages at the modeling level shall be enabled.
		
	\section{Conclusion \& Future Work}\label{conclusion-future-work}
		In this paper, we proposed our vision towards MDE4QAI, where the MDE/MDSE paradigm shall be used to facilitate the transition towards the world of QC and QAI. We advocate enhancing existing general purpose and domain-specific modeling languages, including our prior work for the IoT/CPS domain, ML-Quadrat \cite{ML-Quadrat}, in order to enable a hybrid quantum-classical model for the distributed and heterogeneous software systems, particularly the smart IoT/CPS services/applications of the near future, which shall run on an even more diverse set of underlying technologies, including various quantum processors with different architectures and technologies.

\section*{Acknowledgment}
The authors would like to thank Dalibor Hrg for sharing the insights.

%%% Angabe der .bib-Datei (ohne Endung) / State .bib file (for BibTeX usage)
\bibliography{refs} %\printbibliography if you use biblatex/Biber
\end{document}